\documentclass[prl,aps,showpacs,
nofootinbib,
superscriptaddress,twocolumn,preprintnumbers]{revtex4}

\usepackage{graphicx}

\newcommand{\be}{\begin{equation}}
\newcommand{\ee}{\end{equation}}
\newcommand{\ba}{\begin{eqnarray}}
\newcommand{\ea}{\end{eqnarray}}
\newcommand{\bi}{\begin{itemize}}
\newcommand{\ei}{\end{itemize}}
\newcommand{\bc}{\begin{center}}
\newcommand{\ec}{\end{center}}
\newcommand{\bmin}{\begin{minipage}}
\newcommand{\emin}{\end{minipage}}
\newcommand{\rmi}[1]{{\mbox{\scriptsize #1}}}
\newcommand{\fig}{Fig.~}
\newcommand{\eq}{Eq.~}

\newcommand{\nr}[1]{(\ref{#1})}

\newcommand{\bfx}{{\bf x}}
\newcommand{\bfp}{{\bf p}}
\newcommand{\bfq}{{\bf q}}

\newcommand{\qs}{Q_{\mathrm{s}}}

\newcommand{\gt}{{\gamma^0}}
\newcommand{\gz}{{\gamma^3}}
\newcommand{\half}{\frac{1}{2}}

\newcommand{\ud}{\mathrm{d}}
\newcommand{\xt}{\mathbf{x}_T}

\newcommand{\pt}{{\mathbf{p}_T}}
\newcommand{\qt}{{\mathbf{q}_T}}
\newcommand{\kt}{{\mathbf{k}_T}}

\begin{document}

\title{Chemical thermalization in relativistic heavy ion collisions}

\preprint{HIP-2005-29/TH}
\preprint{SPhT-T05/125}
\preprint{hep-ph/0508229}

\author{F. Gelis} \affiliation{Service de Physique Th\'eorique,
B\^at. 774, CEA/DSM/Saclay, 91191 Gif-sur-Yvette, France}

\author{K. Kajantie} \affiliation{Department of Physics, P.O.Box 64,
FI-00014 University of Helsinki, Finland}

\author{T. Lappi} 
\affiliation{Department of Physics, P.O.Box 64,
FI-00014 University of Helsinki, Finland} \affiliation{Helsinki
Institute of Physics, P.O.Box 64, FI-00014 University of Helsinki,
Finland} 

\begin{abstract}
We compute by numerical integration of the Dirac equation the number
of quark-antiquark pairs initially produced in the classical color
fields of colliding ultrarelativistic nuclei. While the number of
$q\bar q$ pairs is parametrically suppressed in the coupling constant,
we find that in this classical field model their production rate is
comparable to the thermal ratio of gluons/pairs = 9$N_\mathrm{f}$/32. After
isotropization one thus would have a quark-gluon plasma in chemical
equilibrium.
\end{abstract}

\maketitle

Much attention has recently been given to the thermalization
properties of QCD matter formed in ultrarelativistic heavy ion
collisions. Since 1983 \cite{bjorken} hydrodynamic analysis have
assumed rapid initial thermalization and essentially entropy
conserving expansion thereafter.  Now RHIC experiments
\cite{phenixwhitepaper,starwhitepaper}
strongly suggest that this is the correct picture. Theoretically, it
is straightforward to understand the formation of an initial gluonic
system, but the problem has been its isotropization in momentum space
\cite{serreau}.
Weak coupling methods fail if they only include collective effects by
screening \cite{mueller} but show great promise if they include collective effects
caused by the anisotropy of the system 
\cite{stantheman,arnold,strickland,dumitrunara}.
However, this still leaves open the chemical equilibration of
the system \cite{elliottrischke}.

\begin{figure}
\hfill 
\includegraphics[width=0.44\textwidth]{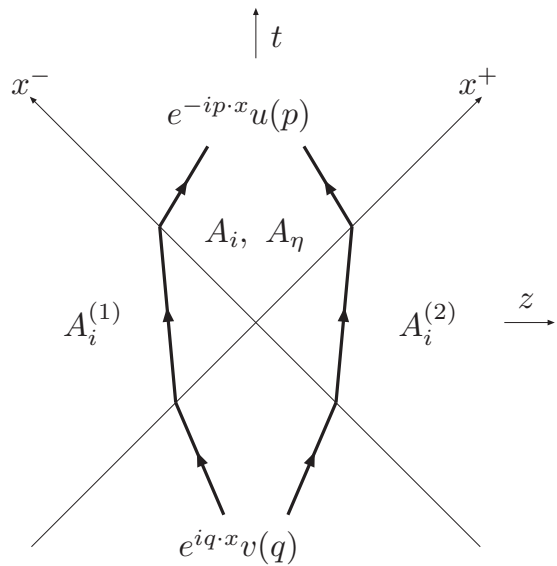}
\hfill
\caption{Domains of different time dependences. The fermion amplitude
is a sum of two terms: one with interaction first with the left moving 
nucleus, then evolution over a region in which the gluon field
$A_i^{(2)}$ is a gauge transform of the vacuum
and finally evolution in a nontrivial gluon field $A_i,A_\eta$.
The other term is symmetric.
\label{tint}}
\end{figure}

The physics involved here is as follows. Two large nuclei of radius
$R_A$ moving along the light cone in opposite directions
(\fig\ref{tint}) collide with zero impact parameter. They correspond
to an ensemble of color currents moving along the light cone in
opposite directions. The stochastic properties of the transverse color
density $\rho^a(\xt)$ have been extensively studied
\cite{iancuvenugopalan,hatta}. The magnitude of the charging is
described by a parameter combination $g\mu$. For given element of the
charge ensemble, color fields can be computed by solving the
Yang-Mills equations $[D_\mu,F^{\mu\nu}]=0$ with initial conditions
given by continuity across the light cone \cite{kmw}. For a dilute
system (small $g^2\mu R_A$) one can compute the gluon fields  analytically
\cite{kmw,kovchegovrischke}.  In general, the color fields can be
obtained by a numerical computation \cite{krasnitzvenu,lappi} and then
interpreted in terms of gluon production. The color fields are
space and time dependent and a definite quantum mechanical probability
of producing $q\bar q$ pairs is associated with them. These are pairs
which are nonperturbatively produced during the first instants of the
collision and which will contribute to chemical equilibration. We will
compute their number and various distributions for numerically
computed gluonic fields.  Again, for weak fields this computation can
be carried out analytically \cite{gelisvenu}.

Assume now that the SU(3) color fields $A_i(\tau,\xt), A_\eta(\tau,\xt)$
\footnote{Our longitudinal variables are $t,z$, $x^\pm = (t \pm
z)/\sqrt{2}=\tau e^{\pm\eta}/\sqrt2$, $\tau = \sqrt{t^2-z^2}=
\sqrt{2x^+ x^-}$, $\eta = \half \ln (x^+/x^-)$, $\ud s^2 = \ud t^2-\ud
z^2=2\ud x^+\ud x^-= \ud \tau^2 - \tau^2 \ud \eta^2$ so that $A_\tau
=A^\tau= (t A^0 - z A^3 )/\tau=(x^+A^-+x^-A^+)/\tau$ and $A_\eta =
-\tau^2 A^\eta = z A^0 - t A^3=x^+A^--x^-A^+$.
We work in the Schwinger gauge $A_\tau=0$.
} are known for
$0\le\tau\lesssim 1/\qs\approx0.2 \textrm{  fm}/c$ (RHIC value). A computation of
pair production requires a careful discussion of in and out vacua and
their relation.  For the present field configuration this has been
carried out in \cite{baltz}. The computation, which is set up so that
one obtains the average number of pairs, which is just the quantity we
want, proceeds as follows.  Choose an antiquark of momentum $\bfq$ and
mass $m$ and solve as a function of time the Dirac equation with this
color field for the spinor $\psi_{\bf q}(t,{\bf x})$ which in the
distant past is given by the negative energy spinor $\psi_{\bf
q}(t\to-\infty,{\bf x})=e^{iq\cdot x}v(q)$.  The time integration
brings in positive energy components and consists of three
qualitatively different domains, see \fig\ref{tint}. The region
$x^\pm<0$ is trivial. The regions marked $A_i^{(1)}, A_i^{(2)}$ can be
dealt with analytically \cite{1+1d} and one obtains an initial condition for
$\psi_\bfq(\tau=0,z,\xt)$ along the positive light cones. This rather
complicated initial condition, given explicitly in \eq(16) of
\cite{1+1d}, depends on the Wilson lines
$U_{(1)}(\xt), U_{(2)}(\xt)$
corresponding to the gauge fields
of the nuclei, the initial color field
$A_i(0,\xt)$ and on $y_q,\qt,z,\xt$. The spinor $\psi_\bfq(t,{\bf x})$
at $\tau>0$ is then computed by solving the Dirac equation in the
given color field forward in time. Finally, one chooses a quark
momentum $\bfp$ and forms the overlap between a positive energy spinor%
\footnote{Whether it is justified to use
a free spinor at a finite $\tau$ in the presence of the external field
merits further study.}
$\phi_\bfp(x)=e^{-ip\cdot x}u(p)$ and the outcome of the time
evolution of the negative energy spinor in the distant past:
\begin{equation}
M_\tau(p,q)\equiv
\int \frac{\tau \ud z \ud^2\xt}{\sqrt{\tau^2+z^2}}
\,\phi_\bfp^\dagger(\tau,\bfx)\gt\gamma^\tau
\psi_\bfq(\tau,\bfx)\; .
\label{mtau}
\end{equation}
The overlap is computed at fixed $\tau$, hence the use of
$\gamma^\tau$, $\gt\gamma^\tau=
\cosh\eta-\gt\gz\sinh\eta=\exp(-\eta\gt\gz)$. This is also the reason
for the Jacobian factor $\tau dz/\sqrt{\tau^2+z^2}$ in the
longitudinal integration. 
We evaluate Eq.~(\ref{mtau}) in the 
2-dimensional Coulomb gauge $\partial_i A_i = 0$. This is the gauge
condition used in the Abelian case \cite{baltz} and also the one
used to evaluate the number of gluons in the background field.
Eq.~(\ref{mtau}) gives us
\begin{equation}
{dN\over dy}=\int
\frac{\ud y_p \ud^2\pt}{ 2 \left( 2\pi \right)^3}
\frac{\ud y_q \ud^2 \qt}{ 2 \left( 2\pi \right)^3}
\delta\bigl(y-y_p )
\left| M_\tau(p,q) \right|^2,
\label{numberofpairs}
\end{equation}
the number of quarks of one flavor of mass $m$ per unit rapidity
(since an equal number of antiquarks
are produced, we refer to this quantity as the ``number of pairs''
below). Since the gluon fields are 
$\eta$--independent, $dN/dy$ is independent of $y$. We shall compute
\nr{numberofpairs} for all $\tau$ but it is only after the 
``formation time''
$\tau\gtrsim 1/\sqrt{q_T^2+m^2}$ that the produced antiquarks can
reinteract. Since one expects $q_T\sim g^2\mu $, this limit
for light quarks is $\tau\gtrsim 1/(g^2\mu)$.

The parameters of the computation are the coupling $g$ (constant in
this semiclassical set-up; we use the phenomenologically relevant
value $g=2$, $\alpha_s=0.3$) the source density parameter $\mu$
(depends on atomic number $A$ and collision energy $\sqrt{s}$) the
nuclear radius $R_A$ and the quark mass $m$ (like with $g$ there is
nothing in this semiclassical set-up which would make $m$ scale
dependent).

Two relevant parameter combinations are $g^2\mu$ and $g^2\mu R_A$. The
first one, $g^2\mu$, is the dominant transverse momentum scale of the
classical background field. It is related to the saturation scale
$\qs$; for $\kt<\qs$ the gluonic system becomes so dense that
nonlinear interactions limit the growth of its density, numerically in
one phenomenological model \cite{ekrt} $\qs\approx0.2\,{\rm
GeV}A^{0.128}\sqrt{s}^{0.19}$, $\qs\approx1$ GeV at RHIC energies and
$\approx2$ GeV at LHC energies.  The dimensionless diluteness
parameter $g^2\mu R_A$ determines the importance of nonlinear strong
field effects.

\begin{figure}
\includegraphics[width=0.43\textwidth]{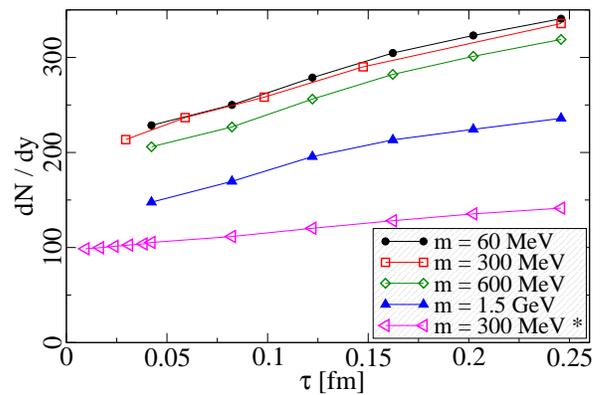}
\caption{
Dependence on proper time $\tau$ of the number of pairs per unit rapidity $\ud N/\ud y$
for $g^2\mu=2$ GeV and for values of quark mass marked on the
figure. The lowest curve corresponds to $g^2\mu=1$ GeV.
\label{inftimelim}}
\end{figure}
\begin{figure}
\includegraphics[width=0.43\textwidth]{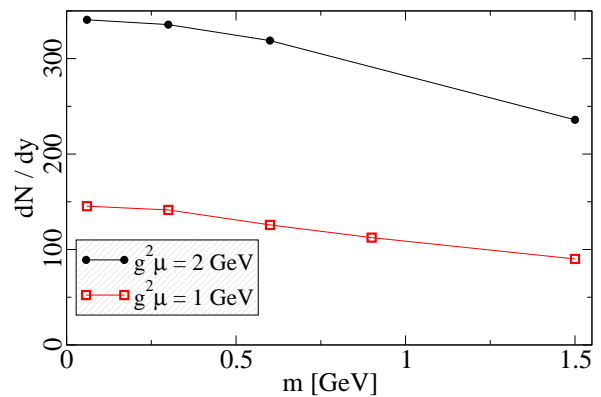}
\caption{Dependence of the number of quark pairs
on quark mass at a fixed proper time, $\tau=0.25 \textrm{ fm}$,
and for two values of $g^2\mu$.
\label{quarkmassdep}}
\end{figure}
\begin{figure}
\includegraphics[width=0.43\textwidth]{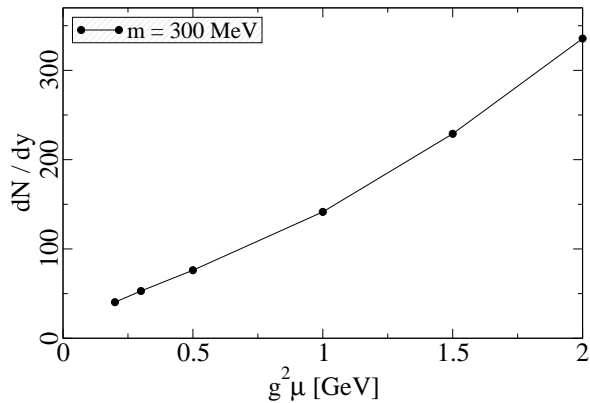}
\caption{Dependence of the number of quark pairs
on $g^2\mu$ at a
fixed proper time, $\tau=0.25 \textrm{ fm}$, and for quark mass $m=0.3 \textrm{ GeV}$.
\label{satscaledep}}
\end{figure}

The numerical computation is done on a $N_T^2N_L$ lattice so that the
total transverse area is $(N_Ta)^2=\pi(6.7\,\textrm{ fm})^2$, i.e., the
transverse lattice spacing is $a=11.8\,{\rm fm}/N_T = 60/N_T \cdot
1/{\rm GeV}$.  The results presented in this letter have been obtained
with $N_T=180, \ N_L=400, \ \ud z = 0.2a$ and $\ud \tau = 0.02a$.  At
each site one has for each color a spinor with 4 complex components,
i.e., ($N_c=3$)$\times 2 \times 4 \times 4=96$ bytes in single
precision, giving a total of $96\cdot180^2\cdot400=1.2 \textrm{ GB}$. This
illustrates the memory requirement of the calculation. The number of
timesteps in the integration in order to reach $\tau = 0.25 \textrm{
fm}$ is of the order of 500.  The numerical method can be tested
by varying these parameters.  Another check of the numerical method is
to study how for a zero external field the result of zero quark pairs
is obtained in a rather nontrivial way by the contributions of the two
paths in Fig.\ref{tint} cancelling each other.

Note that on the transverse lattice one has to use lattice momenta
\begin{equation}
\tilde k^2={4\over a^2}\sum_{i=1}^2\sin^2{ak_i\over2},\quad
\hat k^2={1\over a^2}\sum_{i=1}^2\sin^2{ak_i},
\end{equation}
for bosons and fermions, $ak_i=2\pi n_i/N_T$ with $-N_T/2+1<n_i\le
N_T/2$. 
For fermions the modes $N_T/4 < |n_i| < N_T/2$ are doubler modes, which
we leave out both from the initial condition ($q$ modes) and the projection
to positive energy states ($p$ modes).
For fermions one can effectively only go up to $\hat
k=\sqrt2/a\approx \sqrt{2} N_T/60 \textrm{ GeV} \approx 4 \textrm{
GeV}$ on a $180^2$--lattice.

Results for the number of pairs are shown in
Figs.~\ref{inftimelim}--\ref{satscaledep}.  First,
Fig.~\ref{inftimelim} shows how the pair production amplitude
$M_\tau$ depends on proper
time $\tau$ for $g^2\mu=2$ GeV and for a number of quark mass values,
for $g^2\mu=1$ GeV the $\tau$ dependence for $m=0.3$ GeV is given. A
striking feature of the result is the instantaneous formation of the
pairs at small $\tau$, followed by gradual increase.  To put this in
perspective, note that in the QED case the production amplitude
\nr{mtau} can be computed analytically for all $\tau>0$ and its
square is constant and equal to that of the pair production amplitude
in \cite{baltz}. In Abelian electrodynamics the production thus takes
place instantaneously at $\tau=0$, Here there is also a slow increase
thereafter and one will enter in a domain in which the backreaction of
$q\bar q$ on the gluon fields should be included.  In comparison, for
$g^2\mu=2$ GeV the gluon number grows rapidly to the value $0.29\pi
g^2\mu^2R_A^2\approx1000$ by the time $1/g^2\mu\approx 0.1$ fm
\cite{krasnitzvenu,lappi} and saturates thereafter.

The dependence of the result on quark mass $m$ and on $g^2\mu$ is
shown in Figs.~\ref{quarkmassdep} and \ref{satscaledep}, both at
$\tau=0.25$ fm. The dependence on the quark mass $m$ is surprisingly
weak compared to the perturbative result. 
For numerical reasons we
can not go down to quark masses $m \lesssim 100\textrm{ MeV}$. One should also note
that with the small transverse lattice that we have used here one
cannot study very large masses.  The mass dependence is stronger for
smaller $g^2\mu$, which is expected.

\begin{figure}
\includegraphics[width=0.43\textwidth]{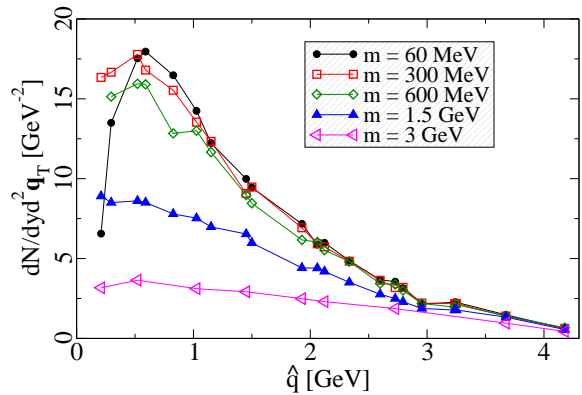}
\caption{Transverse momentum spectrum of (anti)quarks for
$g^2\mu=2$ GeV at a fixed proper time, $\tau=0.25$ fm, and
for different quark masses.
\label{qTgsqmu2}}
\end{figure}
\begin{figure}
\includegraphics[width=0.43\textwidth]{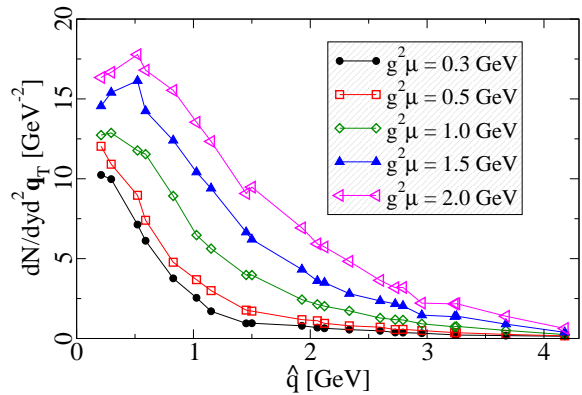}
\caption{Transverse momentum spectrum of (anti)quarks for quark mass
$m=0.3$ GeV and for different $g^2\mu$ at a fixed proper time, $\tau=0.25$ fm.
\label{qTm03}}
\end{figure}

The computation of the gluon fields \cite{krasnitzvenu,lappi} is based
on the Hamiltonian formalism and thus gives directly the transverse
energy $E_T$.  Obtaining the multiplicity is based on assuming free
field dispersion relation. 
In the case of quarks, the pair multiplicity comes directly, and one
must explicitly compute the transverse momentum spectra in order to
obtain $E_T^\rmi{pairs}$.
These are obtained from \nr{numberofpairs} by fixing $y_q=0$
and $\qt$ and integrating over $y_p,\pt$. Results for various quark
masses at fixed $g^2\mu=2$ GeV are shown in Fig.~\ref{qTgsqmu2} and
for various $g^2\mu$ at fixed $m=0.3$ GeV in Fig.~\ref{qTm03}.

One expects $q_T$-spectra to become flatter and smaller with
increasing quark mass $m$ and flatter and larger with increasing
$g^2\mu$. These qualitative features are seen in the result, but
especially the mass dependence is rather weak.

It has conventionally been assumed that the initial state of a
heavy ion collision is dominated by gluons. This is the
result e.g. when both quarks and gluons are produced in
$2 \to 2$ collisions of collinear partons \cite{ekrt}. In the color glass
condensate picture the initial state is entirely gluonic and pair
production is suppressed by a power of $g$.
Our result suggests that quarks could be present in comparable numbers.
This is understandable since the numerical value of the coupling constant $g$ is,
in fact, not small. Note also that the kinematics
are different: the calculation of gluon production in this approach
reduces  in the perturbative limit to a $2 \to 1$ process,
whereas the weak field limit of our present computation is a
$2 \to 2$ process.
To be able to compare quark and gluon production to the same order in $g$
one must compute the first correction to the gluon production result, which 
has not yet been done in the color glass condensate picture. A first step 
in this direction is relaxing the assumption of boost invariance in solving
the gauge field equations of motion \cite{rom}.

The qualitative phenomenological implications of our result are as
follows. Experiments at RHIC have observed about
600 charged ($\approx1000$ charged and neutral) particles per unit rapidity. The
assumption of entropy conservation,
which is supported by the success of ideal hydrodynamics calculations,
then implies as many particles per unit rapidity also in the initial
state, which we are discussing. One has normally
assumed that all the initial particles are gluons. In this framework,
$g^2\mu=2$ GeV would lead to 1000 initial gluons.  Now we see from
\fig\ref{inftimelim} that, for 3 flavors, about 600 pairs would be
produced. This implies that the initial assumption of gluon dominance
should be questioned. In chemical thermal equilibrium the quark/gluon
ratio is $9N_\mathrm{f}/32\approx1$ meaning that 1000 particles in a unit of
rapidity should consist of 330 gluons and 330 pairs. These numbers are
obtained for $g^2\mu \approx 1.2$, i.e., the model gives a consistent
fit with a reduced value of $g^2\mu$, smaller saturation scale.

There is now experimental evidence suggesting that QCD matter formed
in ultrarelativistic heavy ion collisions is from the very beginning
in local kinetic thermal equilibrium. This is what one always has
assumed in applications of hydrodynamics, but its theoretical proof
and understanding has been lacking as the process clearly is
nonperturbative.  We have in this letter shown that the classical
gluon field model with an ensemble of initial conditions also produces
an abundance of quark-antiquark pairs.  In fact, this number is close
to the one dictated by chemical equilibrium between quarks and
gluons. This suggests that the QCD matter formed initially in heavy
ion collisions could be in full chemical and kinetic thermal
equilibrium. Experimentally, this would make it more probable to observe
thermal dilepton radiation in the \mbox{2 GeV} mass range at LHC energies.
It would be very interesting to find also other experimental tests
for initial chemical equilibration.

\acknowledgments{ T.L. was supported by the Finnish Cultural
Foundation.  This research has also been supported by the Academy of
Finland, contract 77744. We thank  R. Venugopalan,
H. Fujii, K. J. Eskola, B. M\"uller and D. Kharzeev for discussions.}

\end{document}